\documentclass[prc,aps,floats,floatfix]{revtex4}
\usepackage{graphicx}
\usepackage{multirow}
\begin{document}

\title{Excitation spectra from angular momentum projection of Hartree-Fock states and 
the configuration-interaction shell-model}

\author{Joshua T. Staker and Calvin W. Johnson}
\affiliation{Department of Physics, San Diego State University,
5500 Campanile Drive, San Diego, CA 92182-1233}

\begin{abstract}
We make numerical comparison of spectra from angular-momentum projection 
on Hartree-Fock states with spectra from configuration-interaction nuclear 
shell-model calculations, all carried out in the same model spaces (in 
this case the $sd$,  lower $pf$, and $p$-$sd_{5/2}$ shells) and 
using the same input Hamiltonians. We find, unsurprisingly, 
that the low-lying excitation spectra for rotational nuclides are well 
reproduced, but the spectra for vibrational nuclides, and more generally the 
complex specta for odd-A and odd-odd nuclides are less 
well reproduced in detail. 
\end{abstract}
\maketitle

\section{Introduction}

In Hartree-Fock (HF) calculations, one approximates the ground state by  a single
Slater determinant (antisymmeterized product of single-particle
wavefunctions) and minimizes the energy, reducing the many-body
problem to an effective one-body problem \cite{ring}. The 
Hartree-Fock solution can break exact symmetries, such as 
rotational invariance, and indeed is 
often \textit{more} effective if it breaks exact symmetries. 
Restoring broken symmetries adds correlations 
that further lower the energy and is a useful approach for many problems. 

In this paper we project states 
of good angular momentum  from a Hartree-Fock state, 
and compare the resulting ground state energies and, especially,  
excitation spectra to exact results from equivalent full configuration-interaction 
diagonalization in a shell-model basis. 
Such comparisons help us to understand the accuracy and limitations of the 
projected Hartree-Fock approximation (PHF), as 
well as potentially providing a useful shortcut to nuclear structure. 

PHF is not new \cite{PY57} and can be found in a number of variants. Most of these 
are inspired by the Nilsson model \cite{Ni55,La80} and the Elliott SU(3) model \cite{El58}, and they 
can be classified by their use of (a) quasiparticle single-particle states (b) mixing of particle-hole 
states and (c) schematic or realistic interactions. For example, the projected 
shell model (PSM)  uses quasiparticle single-particle states but no mixing of particle-hole states and 
schematic interactions \cite{PSM}; the MONSTER and VAMPIR codes use quasiparticle single-particle 
states,  and mixing of
particle-hole states, and realistic interactions \cite{VAMPIRetc}; and the projected 
configuration-interaction (PCI) eschews quasiparticles states but uses realistic 
interactions and mixes higher-order particle-hole states \cite{PCI}. 
This list is not exhaustive. 

What is new in this paper is the comparison of exact numerical solutions 
against a relatively simple state, with a focus on the quality of the excitation spectra.  In this case the ``exact'' solution is 
from a full configuration-interaction (CI) calculation, using semi-realistic 
interactions diagonalized in a truncated but  nontrivial shell-model basis, 
against a single Hartree-Fock state with angular momentum projection. Hence we 
eschew quasiparticle states and particle-hole states, but we use a realistic 
interaction and we allow arbitrary deformation.  

This paper is similar in philosophy to previous work comparing the random phase approximation (RPA) against CI calculations 
in a CI basis \cite{SHERPA,SJ02}, testing \textit{how good is a specific approximation against a 
numerically exact result}; in the former work the approximation was RPA, while here it is PHF.   Another similar set of papers are those on PCI \cite{PCI}, 
which compare results directly to CI diagonalization. 
An important difference with much previous work, however, is that in addition to studying 
the trends of the ground state energies, we consider in detail the excitation spectra.

We find that rotational spectra, and spectra that are from simple particle-rotor coupling, are 
well reproduced, with a small rms error.  Other excitation spectra, such as vibrational spectra or complex spectra
from odd-$A$ and odd-odd nuclides, have larger rms errors. This is unsurprising, but it is useful 
to verify in detail as we have done. 

In the next two sections we outline the numerical exact (CI) and approximate (HF and PHF) 
calculations, which we then compare for a variety of nuclides in Section V 

\section{The shell-model basis and configuration-interaction calculations}

We begin by outlining the shell-model basis in which we work and the fundamentals 
of configuration-interaction calculations \cite{BG77,BW88,SMreview05}. 

We work entirely in occupation or configuration space.  This means 
that we start with some finite set of orthonormal single-particle states, 
$\{ \phi_a(\vec{r}) \}$.  Each of these states have good angular momentum and parity and 
are labeled by orbital $l$, total $j$, and $z$-component $j_z$; implicitly they have 
good spin $s$, which here is 1/2 although other values are allowed by our code.   
We also have two species of particles, here protons and neutrons, each with 
fixed numbers.  

Using second quantization we have fermion creation  and annihiliation 
operators $\hat{\phi}_a^\dagger, \hat{\phi}_a$ that create or remove a particle from 
the states $ \phi_a(\vec{r})$. The uncoupled Hamiltonian is then
\begin{equation}
\hat{H}=\sum_{ab} T_{ab} \hat{\phi}^\dagger_a \hat{\phi}_b 
+ \frac{1}{4} \sum_{abcd} V_{ab,cd}\hat{\phi}^\dagger_a \hat{\phi}^\dagger_b \hat{\phi}_d \hat{\phi}_c 
\label{H2q}
\end{equation}
The one-body and two-body matrix elements, $T_{ab}$ and $V_{ab,cd}$ respectively, 
are appropriate integrals over the Hamiltonian and the single-particle states 
\cite{BG77}.
Because the Hamiltonian is an angular momentum scalar, we store the matrix elements 
in coupled form. Given the single-particle space, the matrix elements are generated 
externally to 
all our codes and read in from a file. Any radial form of the single-particle states and any form of the 
interaction, including non-local interactions, are allowed and make no pratical difference to our 
many-body codes.  Because we work in second quantization, the two-body matrix elements 
are automatically antisymmeteric and 
we do not separate direct from exchange terms. 

For configuration-interaction (CI) calculations we use the BIGSTICK code \cite{BIGSTICK}. 
BIGSTICK creates a many-body basis of Slater determinants $\{ | \alpha \rangle \}$, 
constructed from the single-particle basis:
\begin{equation}
| \alpha \rangle = \prod_{a = 1}^{N_p} \hat{\phi}^\dagger_a | 0 \rangle,
\end{equation}
where $N_p$ is the number of particles (occupied states).  We fix 
the total $M =J_z$ of the many-body basis states, (this is trivial as each single-particle state has good 
$j_z$ as well) and is thus called an $M$-scheme code; because the Hamiltonian is an angular momentum scalar, 
the final eigenstates thus will automatically have good total angular momentum $J$ (and 
isospin $T$, although we have the capability to break isospin; $T_z = (Z-N)/2$ is fixed). 
Within the fixed single-particle space and fixed $J_z, T_z$ we allow all possible configurations.

Given the basis, BIGSTICK then computes the many-body Hamiltonian matrix elements 
$\langle \alpha | \hat{H} | \beta \rangle$ and finds the low-lying eigenstates, including the ground state, using the Lanczos algorithm \cite{Lanczos,SMreview05}.  
With the single-particle space and Hamiltonian matrix elements fixed, the resulting eigenenergies for the 
full-space CI calculation are numerically exact. BIGSTICK can handle CI $M$-scheme model 
spaces on a desktop computer up to dimension roughly 2-400 million.

\section{Hartree-Fock approximation}

The Hartree-Fock approximation (HF) is a variational method using a single Slater determinant
 \cite{ring}. One applies 
a unitary transformation among the single-particle states:
\begin{equation}
\hat{c}_i = \sum_{a=1}^{N_s} U_{ai} \hat{\phi}_a, \label{Usp}
\end{equation}
where $N_s$ is the number of single-particle basis states (hence the number of 
particles $N_p < N_s$) 
and creates the trial Slater determinant 
\begin{equation}
| \Psi_T \rangle =\prod_{i =1}^{N_p} \hat{c}^\dagger_i | 0 \rangle.
\end{equation}
Then one finds a Slater determinant that minimizes the energy, that is, that minimizes
\begin{equation}
E = \frac{ \langle \Psi_T | \hat{H} | \Psi_T \rangle } { \langle \Psi_T |  \Psi_T \rangle }
\end{equation}

Because our input matrix elements are  antisymmeterized, we fully include both direct and exchange 
terms.  Working in occupation space the exchange term causes us no difficulty (or, to put it another way, 
any difficulty is off-shored into calculation of the antisymmeterized integrals). 

Many Hartree-Fock calculations enforce good angular momentum, for closed-shell systems or 
closed-shell plus or minus one particle.  Our HF code (the SHERPA code \cite{SHERPA}) allows the Slater determinant to break rotational 
invariance, even for closed-shell systems; such Slater determinants are `deformed.'  
Experience in nuclear physics suggest this to be a fruitful 
path. Our only constraint is that we assume the transformation (\ref{Usp}) to be real.

In order to minimize, we use the standard Hartree-Fock equations, which consist of 
iteratively solving
\begin{equation}
\mathbf{h}\vec{u}_i = \epsilon_i \vec{u}_i, \label{HF}
\end{equation}
where the Hartree-Fock effective one-body Hamiltonian is 
\begin{equation}
h_{ab} = T_{ab} + \sum_{cd} V_{ac,bd} \rho_{dc}
\end{equation}
and 
\begin{equation}
\rho_{dc} = \sum_{i = 1}^N U_{ic} U_{id}
\end{equation}

We have separate Slater determinants for proton and neutrons; the generalization is straight 
forward. More details are found in  Appendix \ref{HFcomp}. 

\section{Projection of angular momentum and parity}

In order to project out angular momentum, we introduce the 
standard projection operator \cite{La80,ring}
\begin{equation}
\hat{P}^J_{MM^\prime} = 
\int d\Omega {\cal D}^{(J)*}_{MM^\prime} (\Omega) \hat{R}(\Omega),
\label{jproj0}
\end{equation}
where $\Omega = \alpha,\beta,\gamma$ and $d\Omega = d\alpha \sin \beta d\beta d\gamma$,
${\cal D}^{(J)*}_{MM^\prime}$ is the Wigner $D$-matrix \cite{Edmonds}, and 
$\hat{R}(\Omega)$ is the rotation operator.  The rotation operator acts on 
a Slater determinant $U$ whose columns are single-particle states with good 
$j,m$; therefore the matrix elements of $\hat{R}$ are given by the Wigner $D$-matrix:
\begin{equation}
\langle j^\prime m^\prime | \hat{R}| j m \rangle 
= \delta_{j^\prime j} D^j_{m^\prime m}(\alpha \beta \gamma).
\end{equation}
It is useful to note that the $j,m,m^\prime$ for the rotational matrix $\hat{R}$ are those 
of the single-particle space, while the $J,M, M^\prime$ for the projection operator 
(\ref{jproj0}) are those of the many-body space. 


We now introduce the Hamiltonian and ``norm'' matrix elements,
\begin{equation}
h^J_{M,M^\prime} \equiv \langle \Psi|
  \hat{H} \hat{P}^J_{M,M^{\prime}}   | \Psi \rangle , \,\,\,
n^J_{M,M^\prime} \equiv  \langle \Psi|
  \hat{P}^J_{M,M^{\prime}}   | \Psi \rangle 
\end{equation}
These are both Hermitian.

 Then we solve the
generalized eigenvalue equation
\begin{equation} 
\sum_{M^\prime} h^J_{M,M^\prime} g^J_{M^\prime} = E^J \sum_{M^\prime} n^J_{M,M^\prime} g^J_{M^\prime}. \label{PHFdiag}
\end{equation}
where we  allow $g_M$ to be complex.  Although the deformed 
Hartree-Fock state will have an orientation, the final result will be independent of 
orientation; we confirmed this by arbitrarily rotating our HF state. 

PHF will only generate a limited number of states. For a given $J$, the maximum number is 
$(2J+1)$, although fewer can be found if the norm matrix $n$ has zero, or very small, eigenvalues. 
This can happen, for example, if the HF state has axial symmetry.  Therefore there will be 
states missing from the CI spectrum, such as low-lying excited $J=0$ states.

\section{Results}

In this section we apply our calculations to a number of nuclides in the 
$sd$ and $pf$ shells as well as a limited $p$-$sd$ space where we allow for 
parity-projection as well.   We ask a number of questions:

\noindent $\bullet$ Because PHF is a variational theory, the PHF ground state 
energy will be above the exact ground state energy. We ask: what is the \textit{systematics} of that displacement? For example, if the displacement were 
nearly constant, we could use PHF to reliably estimate the true ground state energy.

\noindent $\bullet$ In a similar fashion we ask, how well does PHF yield the excitation 
spectrum? 

\noindent $\bullet$ For odd-$A$ and odd-odd nuclei, how often does PHF yield the 
correct $J^\pi$ assignment for the ground state?

\noindent $\bullet$ For multi-shell calculations, how well does parity+angular-momentum 
projection yield the splitting of positive and negative parity spectra?

Many previous studies have focused on the ground state energy. By contrast, we focus 
on the excitation spectrum, as well as systematics of the ground state energy.

We work with several model spaces and interactions.  All model spaces assume some 
inert core and valence particles; the single-particle states can be thought of as 
spherical harmonic oscillator wavefunctions, but that has no impact on our calculations. 
The interactions we use are all `realistic,' based on effective interactions derived 
from scattering data but with individual matrix elements adjusted to fit many-body data
(see \cite{BG77,br06} for details of methodology).
The three model spaces we work in and their interactions are:

\noindent $\bullet$ \textit{sd}, or the 1$s_{1/2}$-$0d_{3/2}$-$0d_{5/2}$ valence space, 
assuming an inert $^{16}$O core; the interaction is the universal $sd$-interaction `B,'
or USDB \cite{br06};

\noindent $\bullet$ \textit{pf}, or the $1p_{1/2}$-$1p_{3/2}$-$0f_{5/2}$-$0f_{7/2}$ valence space, 
assuming an inert $^{40}$Ca core; the interaction is the monopole-modified Kuo-Brown G-matrix interaction 
version 3G, or KB3G \cite{KB3G};

\noindent $\bullet$ and finally the \textit{psd}, or $0p_{3/2}$-$0p_{1/2}$-$0d_{5/2}$-
$1s_{1/2}$ valence space, assuming an inert $^4$He core; the interation is a 
hybrid of  Cohen-Kurath (CK) matrix elements in
the $0p$ shell\cite{CK65}, the older universal $sd$ interaction of Wildenthal \cite{Wildenthal} in
the $0d_{5/2}$-$1s_{1/2}$ space, and the Millener-Kurath (MK)
$p$-$sd$ cross-shell matrix elements\cite{MK75}.  
We leave out the $0d_{3/2}$ orbit is to make full shell-model
calculations tractable.   Within the $p$ and
$sd$ spaces we use the original spacing of the single-particle
energies for the CK and Wildenthal interactions, respectively, but then
shift the $sd$ single-particle energies up or down relative to the
$p$-shell single particle energies to we get the first $3^-$ state at approximately
$6.1$ MeV above the ground state. The rest of the spectrum, in
particular the first excited $0^+$ state, is not very good, but the
idea is to have a non-trivial model, not exact reproduction of the
spectrum. This model space and interaction allows us to consider model nuclei with 
both parities and to investigate parity-mixing in the HF state.

\begin{figure}
\centering
\includegraphics[scale=0.8]{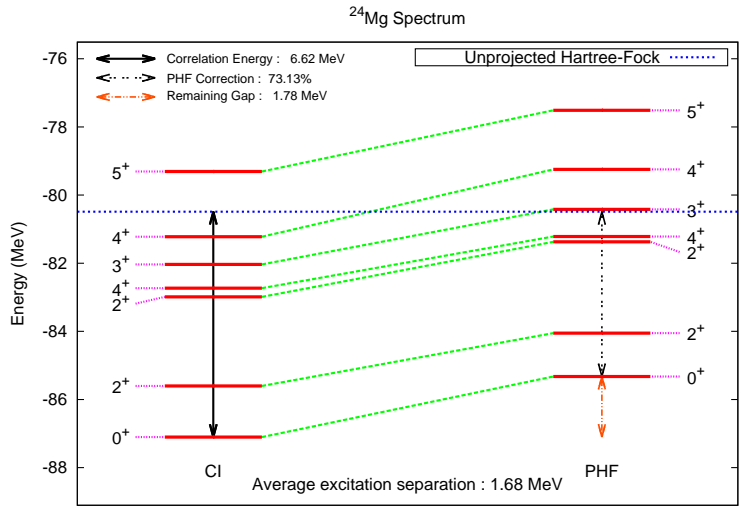}
\caption{Comparison of exact CI and PHF spectra of $^{24}$Mg in the $sd$ model 
space with the USDB interaction. }
\label{Mg24}
\end{figure}

To illustrate our results, we begin with $^{24}$Mg in the $sd$ shell. Fig.~\ref{Mg24}  compares 
the exact CI spectrum on the right against the PHF spectrum on the left.  The PHF g.s. is 
1.78 MeV above the CI g.s., while on average each states in the PHF spectrum is shifted by 
1.68 MeV.  For comparison, the correlation energy, here defined as the difference between 
the unprojected HF energy and the CI g.s. energy, is 6.62 MeV, so that restoring good 
$J$ by projection accounts for $73\%$ of the correlation energy. 

\begin{figure}
\centering
\includegraphics*[scale=0.8]{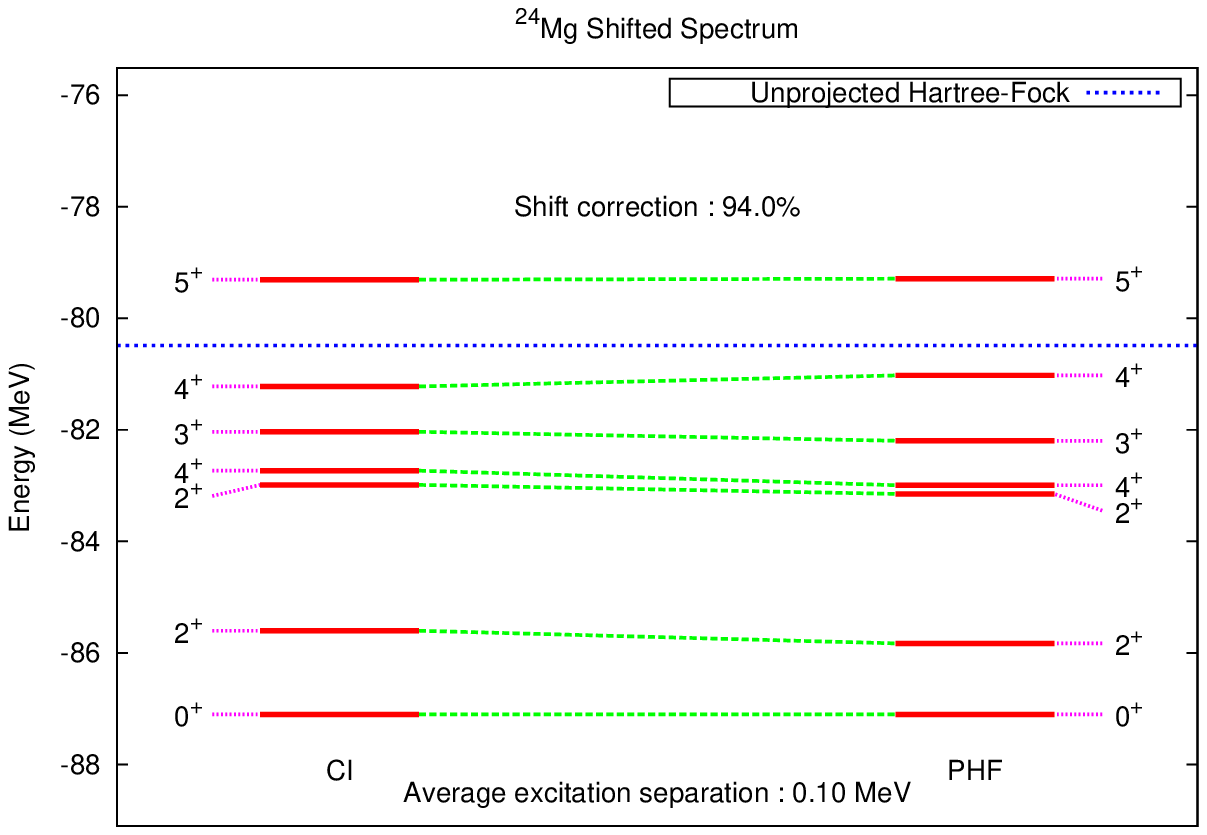}
\caption{Comparison of exact CI and PHF spectra of $^{24}$Mg in the $sd$ model 
space with the USDB interaction, but with the PHF spectrum shifted downwards so the 
g.s. energies coincide. }
\label{Mg24shifted}
\end{figure}

Fig.~\ref{Mg24shifted} is the same as Fig.~\ref{Mg24} but with the PHF spectrum shifted 
down so the g.s. energies coincide.  The agreement between the excitation spectra is very good, 
with an rms error of 0.10 MeV. 

Other nuclides shall qualitatively similar results, albeit with varying degrees of accuracy. In 
general the PHF excitation spectra of even-even nuclides was of much higher accuracy than 
for odd-odd and odd-$A$ nuclides, which is not too surprising as one might expect a 
deformed HF state to approximate a rotational band. 

\begin{figure}
\centering
\includegraphics[scale=0.8]{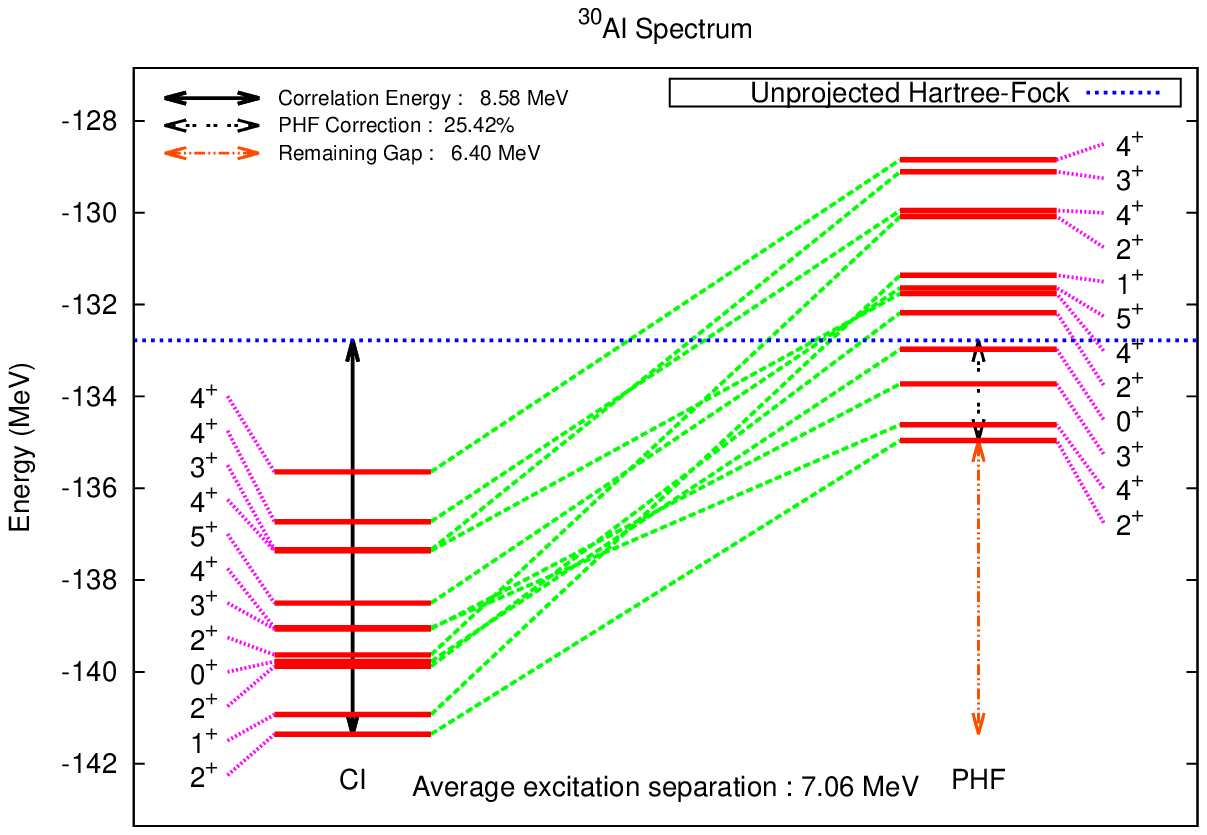}
\caption{Comparison of exact CI and PHF spectra of $^{30}$Al in the $sd$ model 
space with the USDB interaction. }
\label{Al30}
\end{figure}

\begin{figure}
\centering
\includegraphics*[scale=0.8]{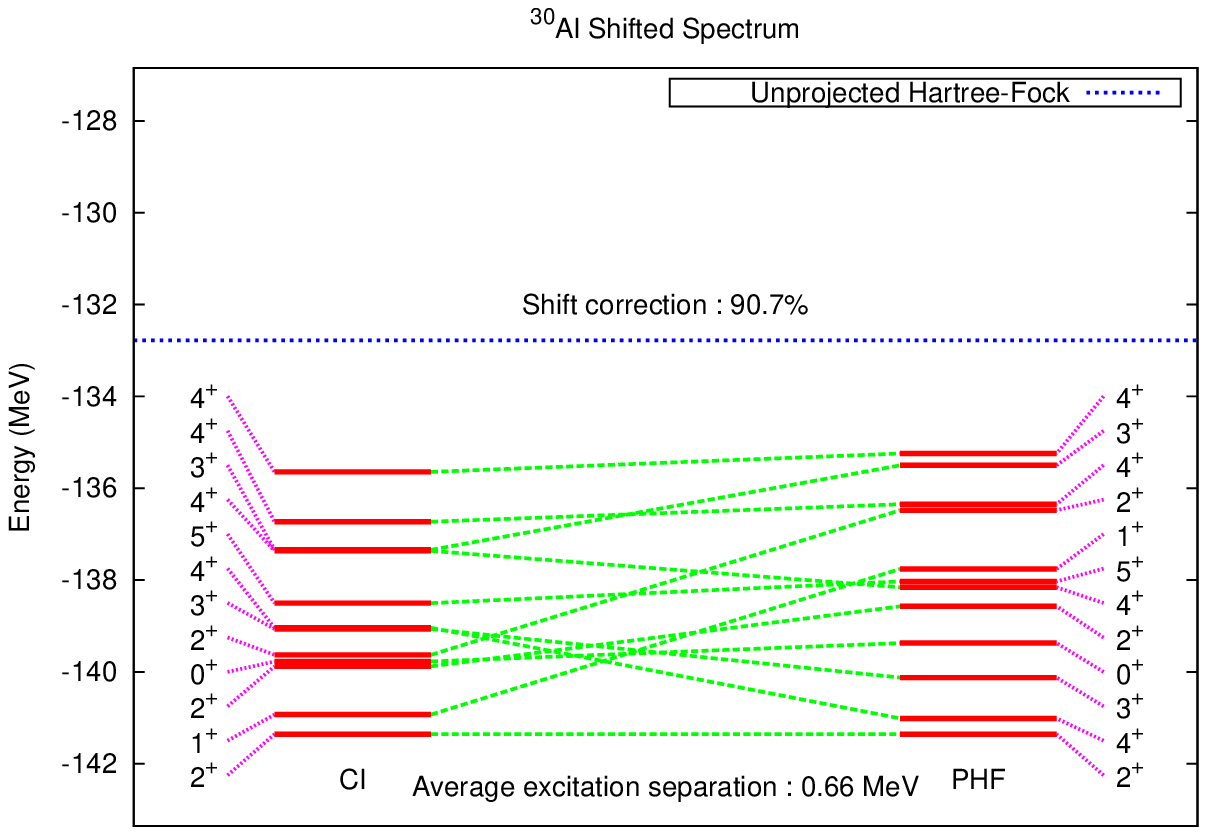}
\caption{Comparison of exact CI and PHF spectra of $^{30}$Al in the $sd$ model 
space with the USDB interaction, but with the PHF spectrum shifted downwards so the 
g.s. energies coincide. }
\label{Al30shifted}
\end{figure}

For example, consider the odd-odd nuclide $^{30}$Al in the $sd$ shell, as shown in 
Fig.~\ref{Al30}.   The PHF g.s. is 6.40 MeV above the CI g.s., and PHF only accounts for 
$25\%$ of the correlation energy.   When the PHF g.s. is shifted down to the match the CI g.s., 
Fig.~\ref{Al30shifted}, the rms error in the excitation spectrum is 0.66 MeV and many of the states are out of order--in particular the first excited  $1^+$ state. 

\begin{figure}
\centering
\includegraphics[scale=0.8]{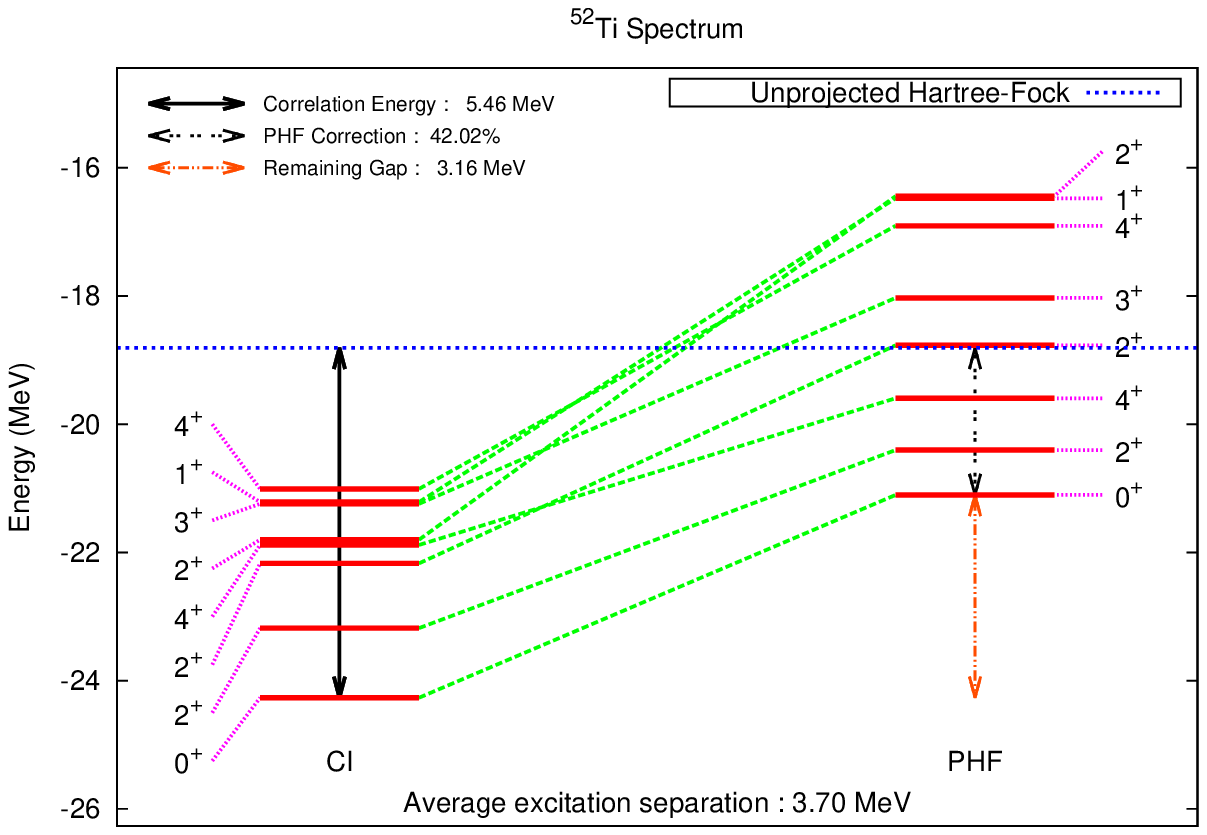}
\caption{Comparison of exact CI and PHF spectra of $^{52}$Ti in the $pf$ model 
space with the KB3G interaction. }
\label{Ti52}
\end{figure}

\begin{figure}
\centering
\includegraphics*[scale=0.8]{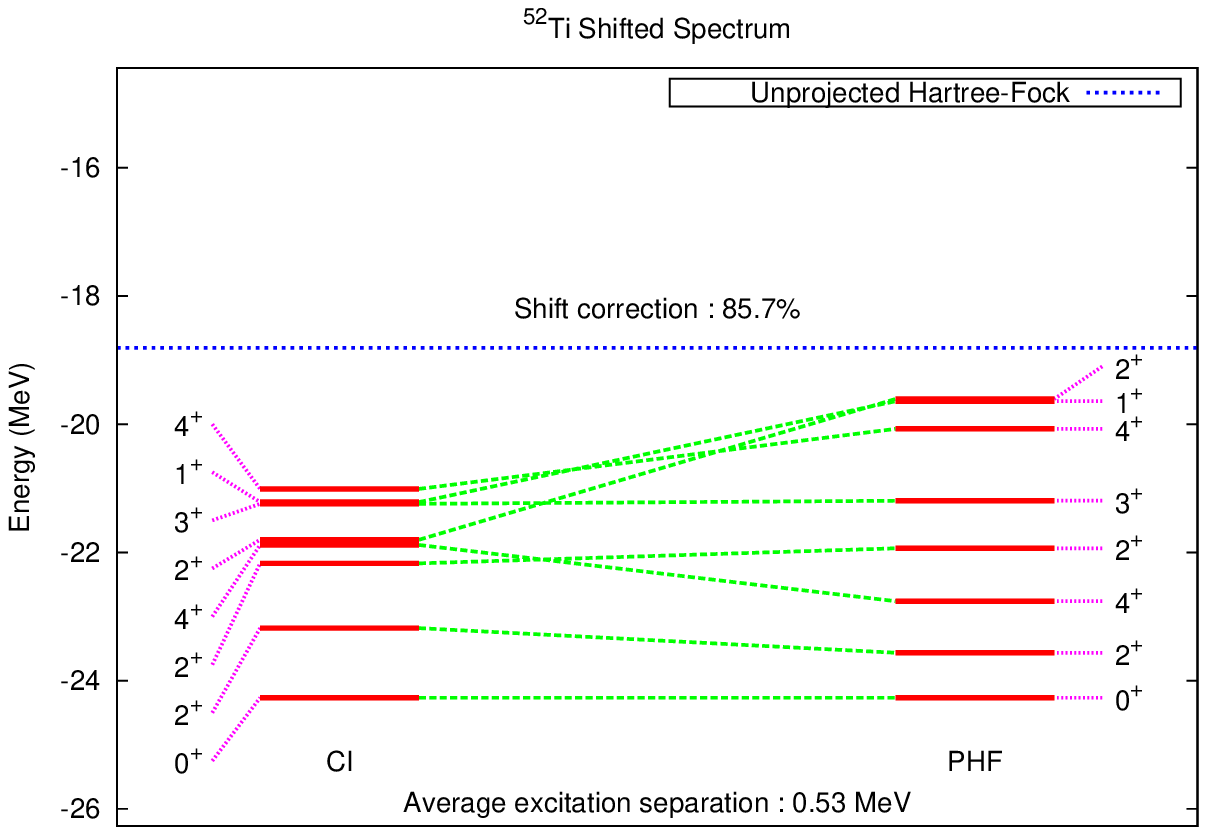}
\caption{Comparison of exact CI and PHF spectra of $^{52}$Ti in the $pf$ model 
space with the KB3G interaction, but with the PHF spectrum shifted downwards so the 
g.s. energies coincide. }
\label{Ti52shifted}
\end{figure}

This continues in the $pf$ shell.  In Fig.~\ref{Ti52} we show $^{52}$Ti, which has a 
strong vibrational rather than rotational spectrum.   Here the PHF g.s. energy is 3.16 MeV above the
CI g.s., account for only $42\%$ of the correlation energy, and even when the PHF spectrum 
is shifted down in Fig.~\ref{Ti52shifted}, 
it is clearly compressed relative to the exact CI, with an rms error of 0.53 MeV. 

\begin{figure}
\centering
\includegraphics[scale=0.8]{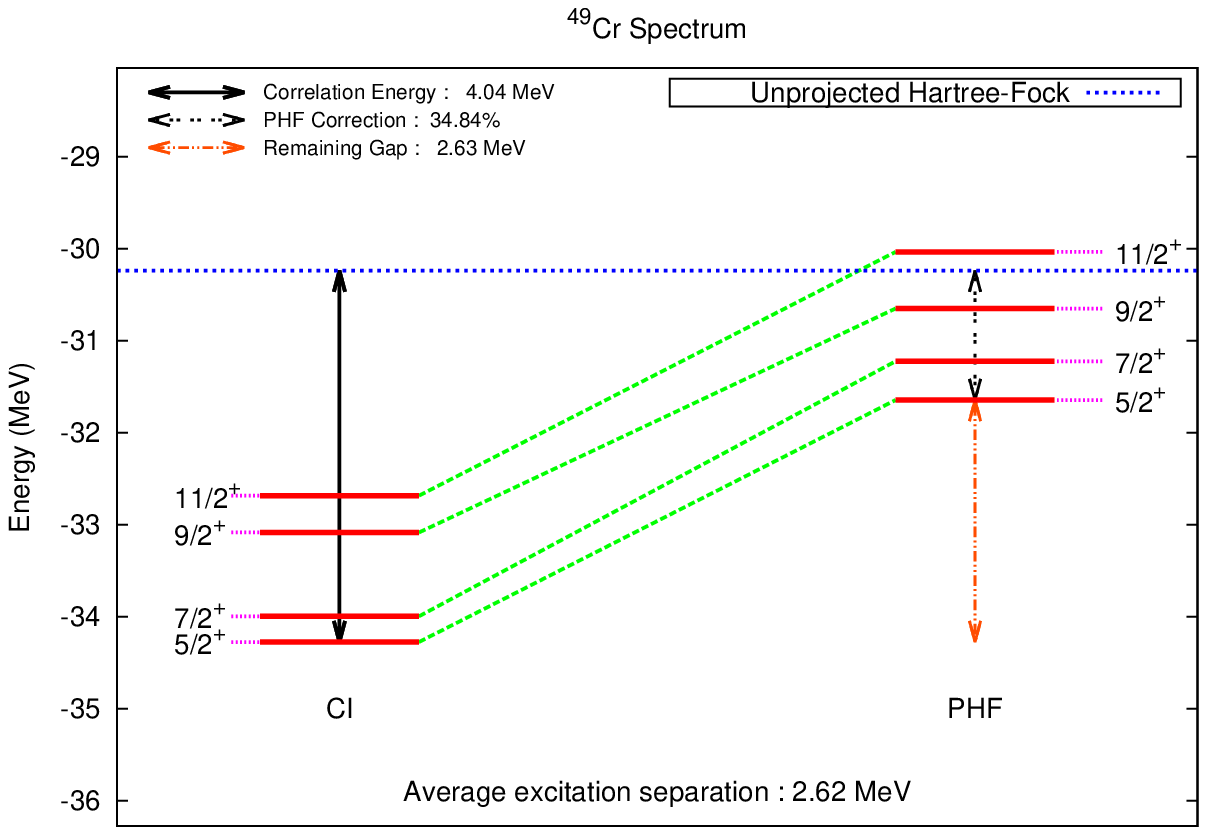}
\caption{Comparison of exact CI and PHF spectra of $^{49}$Cr in the $pf$ model 
space with the KB3G interaction. }
\label{Cr49}
\end{figure}

\begin{figure}
\centering
\includegraphics*[scale=0.8]{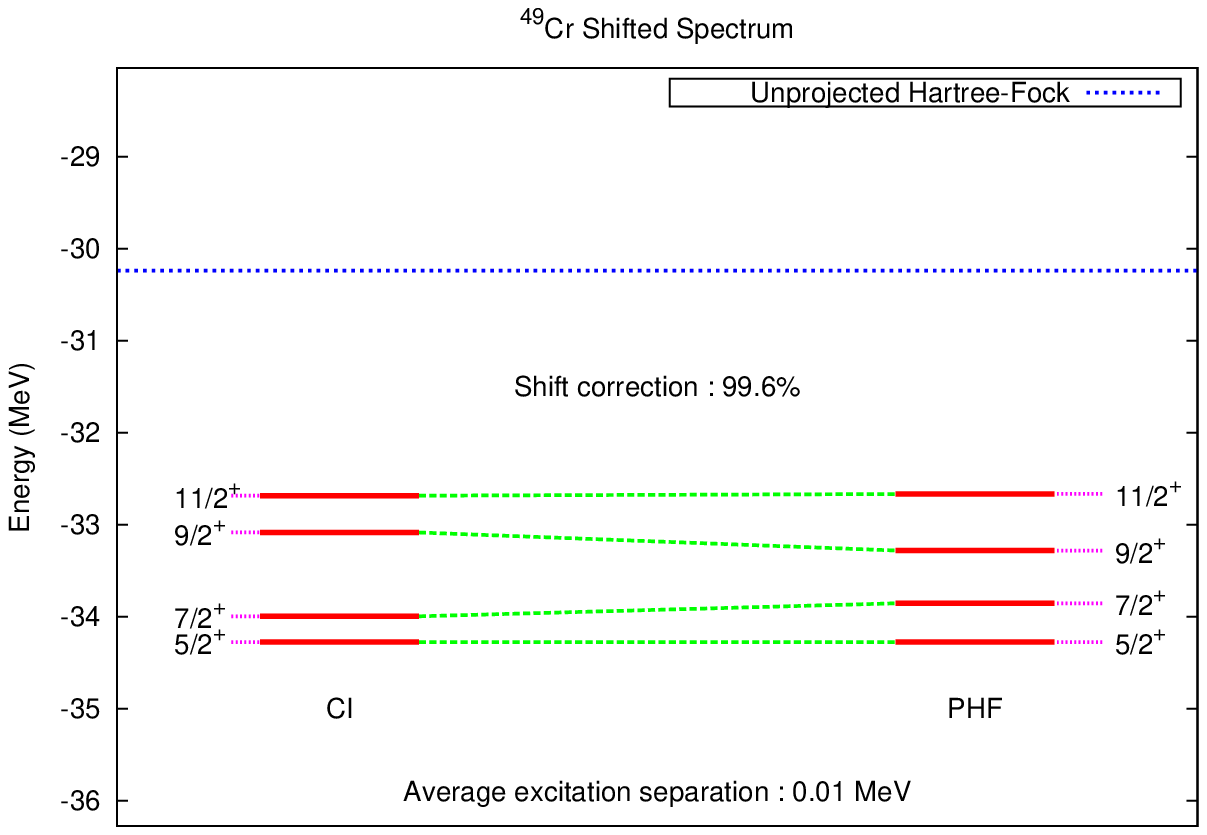}
\caption{Comparison of exact CI and PHF spectra of $^{49}$Cr in the $pf$ model 
space with the KB3G interaction, but with the PHF spectrum shifted downwards so the 
g.s. energies coincide. }
\label{Cr49shifted}
\end{figure}

By contrast, the spectrum of $^{49}$Cr, which can be thought of as a particle-rotor coupled 
nucleus, is good. As shown in Fig. ~\ref{Cr49}, the PHF g.s. is 2.63 MeV above the CI g.s., and PHF
accounts for only $35\%$ of the correlation energy; but when the PHF spectrum is shifted down, 
as in Fig.~\ref{Cr49shifted}, the rms error is only 0.01 MeV.  

We also worked in a cross-shell system, the $p$-$sd_{5/2}$ space.  Keep in mind 
we artificially reduced the separation between the $p$ and $sd_{5/2}$ shells in order to 
provoke a parity-mixed HF state.

\begin{figure}
\centering
\includegraphics[scale=0.8]{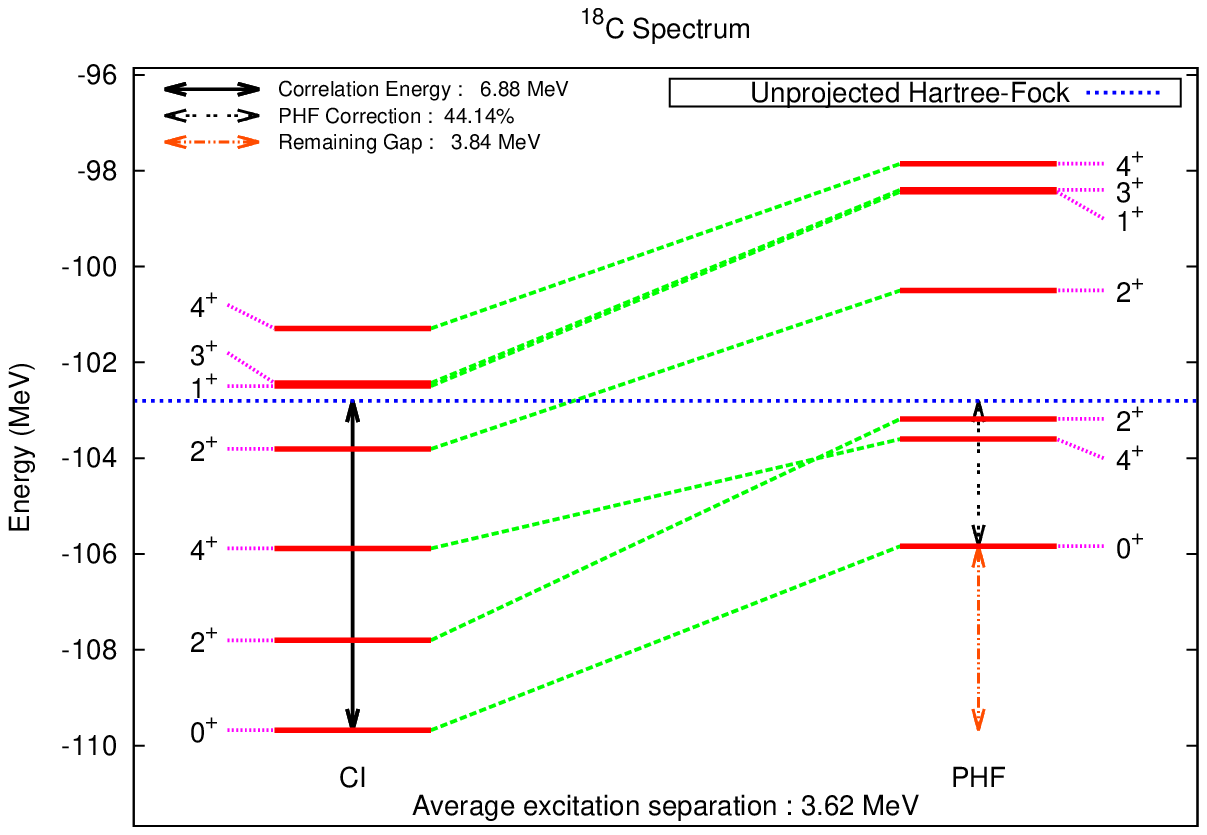}
\caption{Comparison of exact CI and PHF spectra of $^{18}$C in the $p$-$sd_{5/2}$ model 
space. }
\label{C18}
\end{figure}

\begin{figure}
\centering
\includegraphics*[scale=0.8]{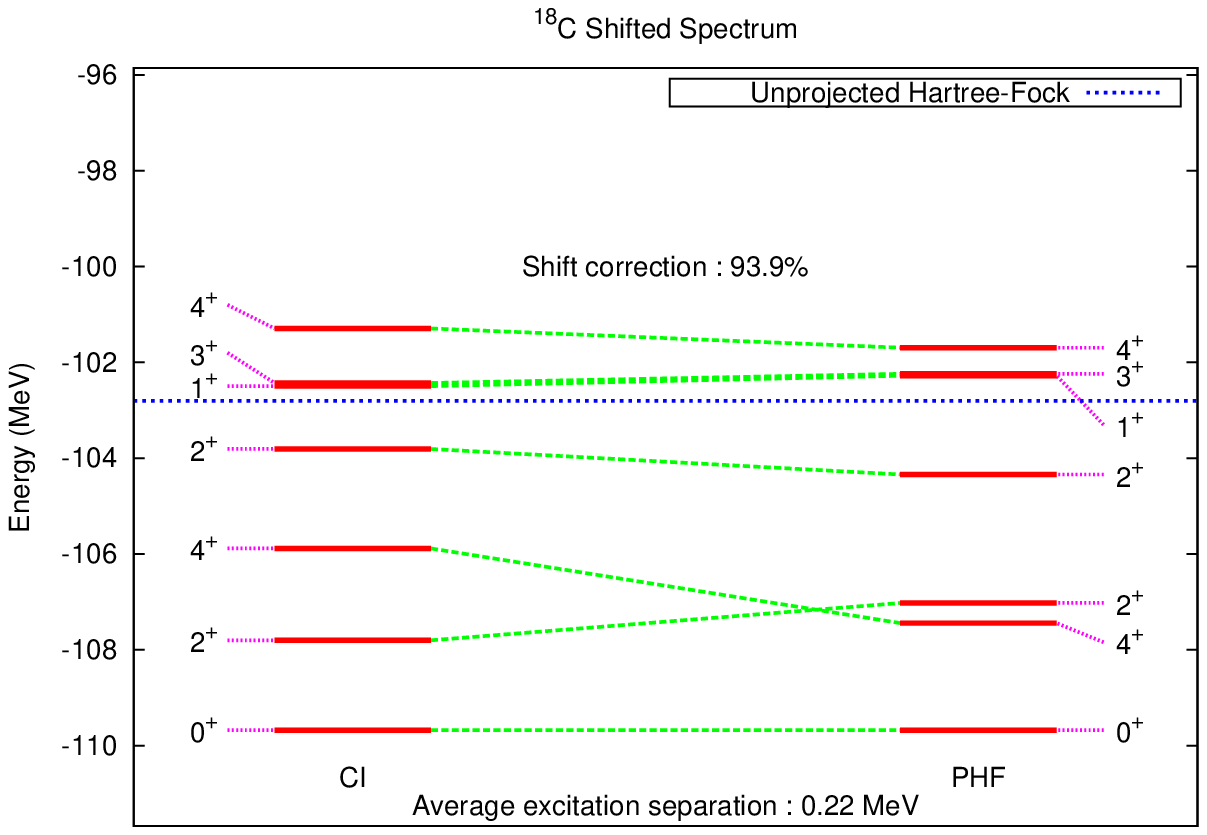}
\caption{Comparison of exact CI and PHF spectra of $^{18}$C in the  $p$-$sd_{5/2}$ model 
space, but with the PHF spectrum shifted downwards so the 
g.s. energies coincide. }
\label{C18shifted}
\end{figure}

 The vibrational spectrum the positive parity states 
of $^{18}$C  is only modestly reproduced by PHF, 
as illustrated in Fig.~\ref{C18} and \ref{C18shifted}.  The PHF g.s. is  3.84 MeV above the 
CI g.s., accounting for only $44\%$ of the correlation energy, and when the PHF spectrum is 
shifted to match the g.s. energies, the rms error is 0.22 MeV, with the $4^+$ state in the
wrong place. 

\begin{figure}
\centering
\includegraphics[scale=0.8]{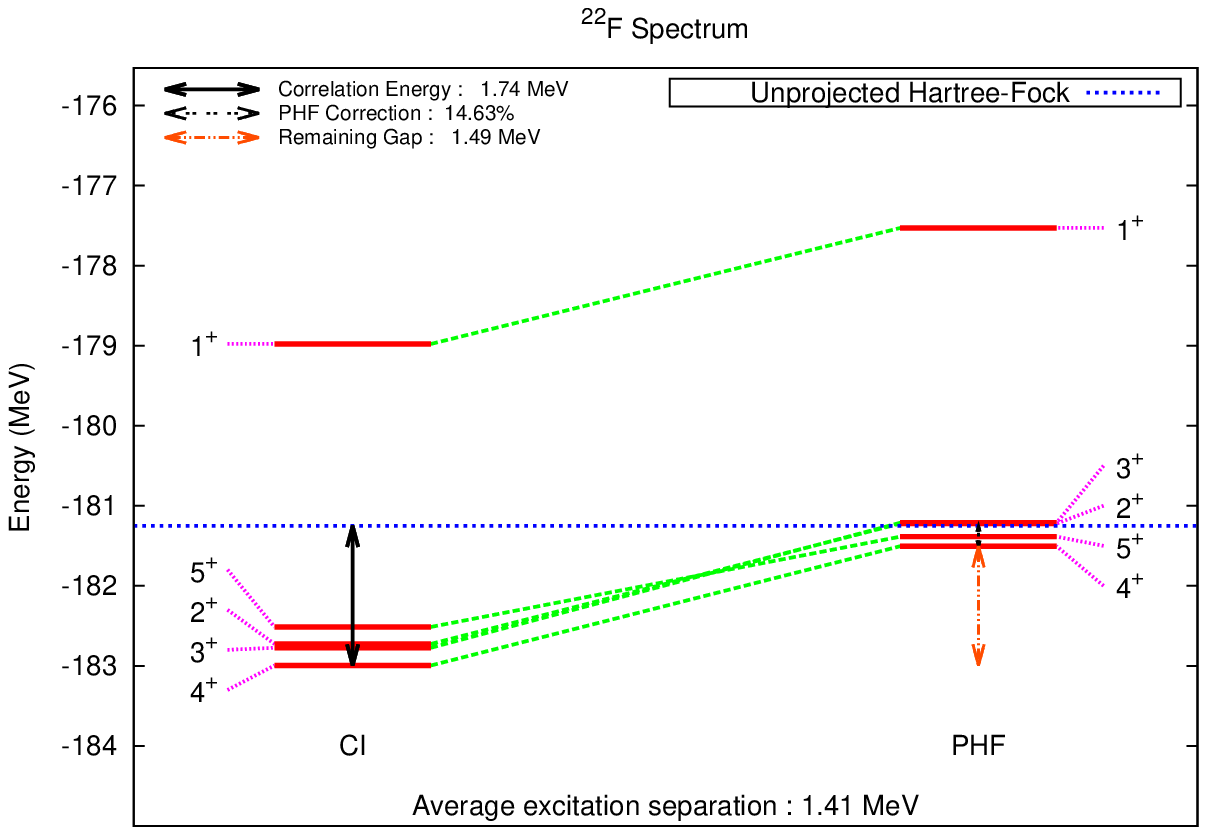}
\caption{Comparison of exact CI and PHF spectra of $^{22}$F in the $p$-$sd_{5/2}$ model 
space. }
\label{F22}
\end{figure}

\begin{figure}
\centering
\includegraphics*[scale=0.8]{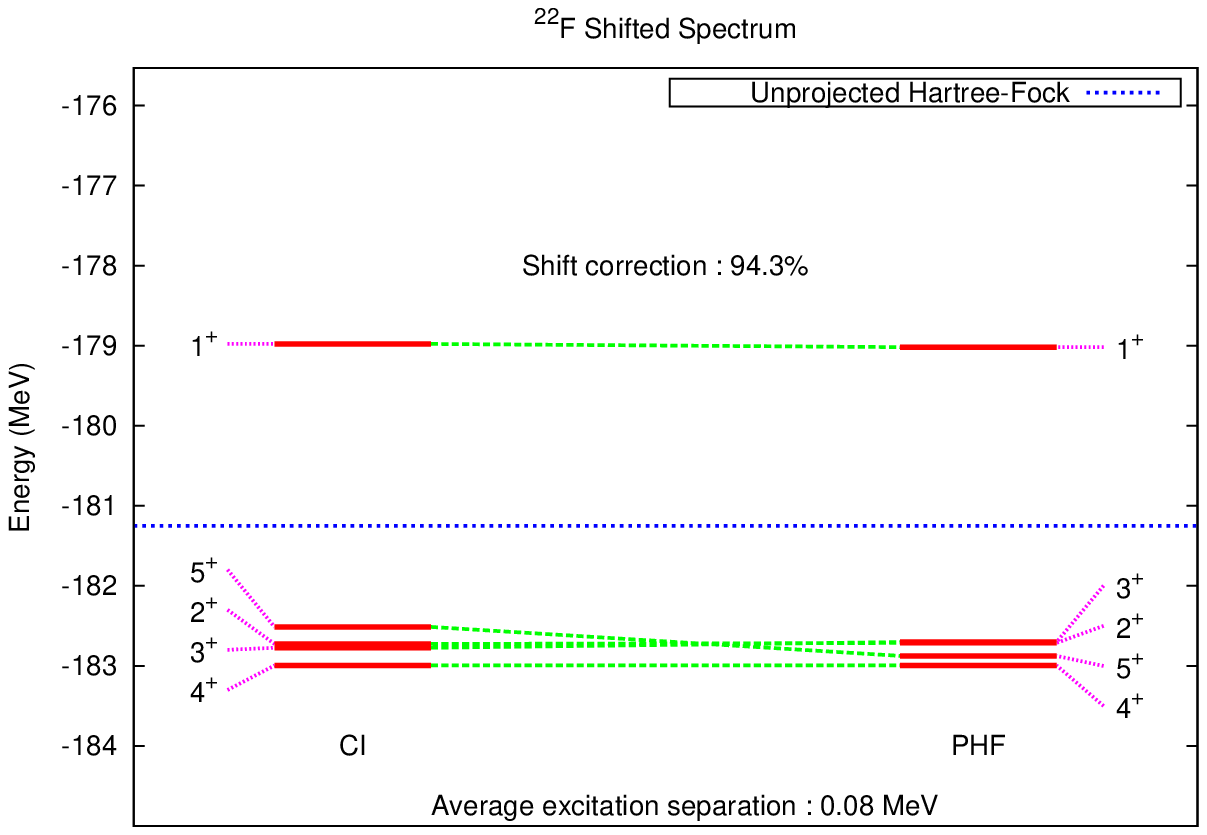}
\caption{Comparison of exact CI and PHF spectra of $^{22}$F in the  $p$-$sd_{5/2}$ model 
space, but with the PHF spectrum shifted downwards so the 
g.s. energies coincide. }
\label{F22shifted}

\end{figure}

Somewhat better are the positive parity states of $^{22}$F, in Figs.~\ref{F22} and \ref{F22shifted}. 
The PHF state is 1.41 MeV above the CI g.s., but that only accounts for barely  $15\%$ of the 
correlation energy. On the other hand, when the g.s. energies are matched, the error in the excitation
is only 0.08 MeV, although some of the states are in the wrong order.

Tables I and II summarize our results.  In particular, Table I summarizes for specific cases 
the standard deviation (rms error) of the PHF excitation spectra relative to the CI excitation spectra, 
once the ground state energies are matched.  Table II summarizes all our calculations; the 
rightmost column gives how often the correct g.s. $J^\pi$ arises out of the PHF.  Even-even nuclides 
almost always get the correct $J=0$ ground state (but for our mixed-shell calculations, occasionally 
the wrong parity--but keep in mind we artificially reduced the splitting between the shells to 
force a mixed-parity HF state), but odd-$A$ got the correct g.s. assignment only a little over 
half the time, and odd-odd nuclei less than half the time. 

\begin{table}
	\centering
	\parbox{5in}{\caption{Summary of data for the nuclei in the paper.  The standard 
deviation is for the shifted PHF spectra relative to the exact CI spectra.\newline}}
	\begin{tabular}{|c|c|cc|c|c|}
\hline
		&&\multicolumn{2}{c}{\textbf{Valence}}&\textbf{Std. dev.}&\textbf{$\%$ PHF} \\

		\textbf{Shell}&\textbf{Nucleus}&\textbf{Z}&\textbf{N}  &(MeV) &\textbf{states} \\ 
\hline
		\multirow{2}{*}{$sd$}&$^{24}$Mg&4&4&0.077&7 \\ 
		&$^{30}$Al&5&9  &1.004&12 \\  \hline
		\multirow{2}{*}{$pf$}&$^{52}$Ti&2&10 &0.714&8 \\ 
		&$^{49}$Cr&4&5 &0.074&4 \\  \hline
		\multirow{2}{*}{$p$-$sd_{5/2}$}&$^{18}$C&4&10 &0.810&7\\ 
		 &$^{22}$F&7&11&0.194&5 \\ \hline
\hline
	\end{tabular}
\end{table}

\begin{table}
	\centering
	\parbox{4.75in}{\caption{Frequency of ground state matching. The ``No. nuclei" column represents the total number of nuclei that was studied for that shell that satisfies the Z-N relationship. \newline}}
	\begin{tabular}{|cccccc|}
	\hline
		&&\textbf{No.}&\textbf{Correct}&\textbf{Incorrect}&\textbf{Frequency} \\
		\textbf{Z - N}&\textbf{Shell}&\textbf{nuclei}&\textbf{G.S.} $\mathbf{J}$&\textbf{G.S.} $\mathbf{J}$&\textbf{correct} \\ \hline
		\multirow{3}{*}{Even-Even}&$s$-$d$&15&15&0&1.00 \\
		&$p$-$f$&10&10&0&1.00 \\
		&$p$-$sd$&17&12&5&0.71 \\ \hline
		\multirow{3}{*}{Odd-Odd}&$s$-$d$&15&8&7&0.53 \\
		&$p$-$f$&6&2&4&0.33 \\
		&$p$-$sd$&12&4&8&0.33 \\ \hline
		\multirow{3}{*}{Odd-A}&$s$-$d$&25&17&8&0.68 \\
		&$p$-$f$&11&6&5&0.55 \\
		&$p$-$sd$&25&13&12&0.52 \\
	\hline
	\end{tabular}
\end{table}

\section{Conclusions}

We have carried out detailed angular momentum projected Hartree-Fock calculations  in a 
shell-model basis with semi-realistic shell-model interactions, and 
compared the spectra to exact configuration-interaction diagonalization calculations, 
with a particular focus on the quality of the excitation spectra.  We found, unsurprisingly, 
that rotational nuclei are best reproduced by PHF.

The U.S.~Department of Energy supported this investigation through
grant DE-FG02-96ER40985.

\appendix
\section{Some computational details for Hartree-Fock}
\label{HFcomp}

A single general Slater determinant we write as a rectangular, 
$N_s \times N_p$ matrix $\mathbf{\Psi}$.  Operationally, the vectors 
$\vec{u}_i$, of length $N_s$, from
solving the Hartree-Fock equation (\ref{HF}) form the columns of 
the transformation of the single-particle basis (\ref{Usp}). We 
take the $N_p$ vectors with lowest $\epsilon_\lambda$ to form the 
columns of $\mathbf{\Psi}$. In other words, $\mathbf{\Psi}$ is simply 
selecting columns of $\mathbf{U}$, which are computed in Eq.~(\ref{HF}); this 
is the unitary transformation that diagonalizes $\mathbf{h}$ . We have separate proton and neutron 
Slater determinants, $\mathbf{\Psi}_\pi$ and $\mathbf{\Psi}_\nu$.

From the Slater determinant we construct the one-body density matrix, 
\begin{equation}
\rho_{ab} = \left \langle \Psi \left | \hat{\phi}^\dagger_a \hat{\phi}_b 
\right | \Psi \right \rangle = \sum_{i = 1}^{N_p} \Psi_{ai} \Psi^*_{bi}.
\end{equation}
We then construct the Hartree-Fock effective one-body Hamiltonian, 
\begin{equation}
h_{ab} = H^1_{ab} + \sum_{cd} V_{ac,bd}\rho_{cd}.
\end{equation}

Because we have two species, we have to take a little care. If our interaction is 
an isospin scalar, then 
\begin{equation}
V^{pp}_{abcd} =V^{nn}_{abcd} = V^{T=1}_{abcd} 
\end{equation}
and
\begin{equation}
V^{pn}_{abcd} =\frac{1}{2}\left( V^{T=0}_{abcd} + V^{T=1}_{abcd} \right).
\end{equation}
Here we have suppressed the obvious decoupling of angular momentum \cite{BG77,Edmonds}.

Then the proton Hartree-Fock Hamiltonian is
\begin{equation}
h^\pi_{ab} = H^1_{ab} + \sum_{cd} V^{pp}_{ac,bd}\rho^\pi_{cd} + 
 V^{pn}_{ac,bd}\rho^\nu_{cd}
\end{equation}
and similarly for the neutron Hartree-Fock Hamiltonian. 

\section{Some computational details for projected Hartree-Fock}

We now have to compute $\langle \Psi | \hat{R}(\Omega) | \psi \rangle$ and 
$\langle \Psi | \hat{H} \hat{R}(\Omega) | \psi \rangle$, where $\hat{R}$ is 
the rotation operator.  In our case
\begin{equation}
\hat{R}| \Psi \rangle \rightarrow \mathbf{R} \mathbf{\Psi} = \tilde{\mathbf{\Psi}}
\end{equation}
where $\mathbf{\Psi}$ and $\tilde{\mathbf{\Psi}}$ are $N_s \times N_p$ matrices 
representing our original and rotated Slater determinants, respectively, and 
$\mathbf{R}$ is the square matrix the carries out the rotation.  In our single-particle 
basis the matrix elements are straightforward:
\begin{equation}
R_{ab} = \langle j_a m_a | \hat{R}(\alpha, \beta, \gamma) | j_b m_m \rangle
= \delta_{j_a j_b} {\cal D}^{j_a}_{m_a, m_b}(\alpha, \beta, \gamma),
\end{equation}
where ${\cal D}^{j}_{m^\prime, m}$ is the Wigner $D$-matrix \cite{Edmonds}.

Compute the matrix elements between two oblique Slater determinants is 
relatively straightforward \cite{Lang93}. First, 
\begin{equation}
\langle \Psi | \hat{R}|{\Psi} \rangle =
\langle \Psi | \tilde{\Psi} \rangle = \det \Psi^\dagger \tilde{\Psi}.
\end{equation}
Calculation of the Hamiltonian (and other) expectation value requires 
the density matrix, 
\begin{equation}
\rho^\prime_{ab} = \frac{ \langle \Psi | \hat{\phi}^\dagger_a 
\hat{\phi}_b|\tilde{\Psi} \rangle }
{ \langle \Psi | \tilde{\Psi} \rangle} 
= \left [ 
\tilde{\mathbf{\Psi}} \left(  \mathbf{\Psi}^\dagger \tilde{\mathbf{\Psi}}\right )^{-1} 
\mathbf{\Psi}^\dagger \right ]_{ba}
\end{equation}
Then 
\begin{equation}
\langle \Psi | \hat{H} | \tilde{\Psi} \rangle 
= \sum_{ab} H^1_{ab} \rho^\prime_{ab} 
+ \sum_{abcd}V_{ab,cd}( \rho^\prime_{ac} \rho^\prime_{bd} - \rho^\prime_{ad} \rho^\prime_{bc} )
\end{equation}
Accounting for two species is straightforward.
Note that even if our original Slater determinants were real, by rotating over all 
angle they can become complex. 

We only project out good angular momentum. We could in principle project out 
exact isospin; to do this we would have to use a single Slater determinant containing 
both protons and neutrons and rotate in isospin space. We leave such a modification for 
future work.

\section{Linear algebra of projected states}

In order to calculate the weights of the angular momentum-projected states in the Hartree-Fock state, we
have to investigate in some detail the linear algebra of solving Eq (\ref{PHFdiag}).

Specifically, we want to expand the Hartree-Fock state
\begin{equation}
| HF \rangle = \sum_\lambda a_\lambda | \lambda \rangle.
\end{equation}
Now each state $\lambda$ has definite $J$, but can be broken up into 
$M$-components. Thus  we expand
\begin{equation}
| HF \rangle = \sum_{\lambda,M}  a_{\lambda, M} | \lambda, M \rangle
\end{equation}
and we ultimately want 
\begin{equation}
|a_\lambda|^2 = \sum_M |a_{\lambda, M}|^2.
\end{equation}

As discussed above, we want to solve the generalized eigenvalue problem
\begin{equation}
\mathbf{\cal H} \vec{g}_\lambda = E_\lambda \mathbf{ \cal N} \vec{g}_\lambda
\end{equation}

Now because 
\begin{equation}
\sum{J KM} \hat{P}^{J}_{K,M} = 1
\end{equation}
we have $\mathrm{tr} \, \mathbf{\cal N} = 1$. 

To solve the generalized eigenvalue problem we decompose the norm matrix, 
$\mathbf{\cal N} = \mathbf{\cal S S}^\dagger$, for example by Cholesky or 
by diagonalization. In general there will be a null space which we project out. 

By creating $\mathbf{h} = \mathbf{\cal S}^{-1} \mathbf{\cal H} \mathbf{\cal S}^{-1 \dagger}$
(and also projecting out the null space, to eliminate division by zero), we then solve the 
ordinary eigenvalue problem
\begin{equation}
\mathbf{h} \vec{v}_\lambda = E_\lambda \vec{v}_\lambda.
\end{equation}
Of course, $| \vec{v}_\lambda |^2 = 1$.  Furthermore, because of this, the vectors $\vec{g}_\lambda 
= mathbf{\cal S}^{-1 \dagger} \vec{v}_\lambda$ are orthonormal with respect to the norm matrix, that is
\begin{equation}
\vec{g}^\dagger_\lambda \mathbf{\cal N} \vec{g}_\mu = \delta_{\lambda \mu}.
\label{normofgvec}
\end{equation}

Another way to see this is to note that 
\begin{equation}
| \lambda, J M \rangle = \sum_K \hat{P}^J_{M,K} | HF \rangle g_K(\lambda)
\end{equation}
and because $\hat{P}^2 = \hat{P}$ we automatically get (\ref{normofgvec}).

As stated above, our ultimate goal is to compute $ |a_\lambda|^2$.  Towards this end, 
we note that 
\begin{equation}
a_{\lambda, M} = \langle \lambda, JM| HF \rangle 
=  \sum_K {\cal N}^J_{M,K} g_K(\lambda) 
\end{equation}
so that 
\begin{equation}
|a_\lambda|^2 = \sum_M |a_{\lambda, M}|^2 
= \sum_{K, K^\prime} g^*_{K}(\lambda) ( {\cal N}^2)^2_{K,K^\prime} g_{K^\prime}(\lambda)
= \vec{g}_\lambda^\dagger \mathbf{\cal N}^2 \vec{g}_\lambda.
\end{equation}
We can rewrite this as 
\begin{equation}
| a_\lambda|^2 = \vec{v}_\lambda^\dagger \mathbf{\cal S}^\dagger \mathbf{\cal S} \vec{v}_\lambda
\label{weight}
\end{equation}
which is what we want in convenient terms. To confirm this is correct, we compute
\begin{equation}
\sum_\lambda |a_\lambda|^2 = \sum_\lambda \vec{v}_\lambda^\dagger \mathbf{\cal S}^\dagger \mathbf{\cal S} \vec{v}_\lambda
= \mathrm{ tr} \, \left( \mathbf{\cal S} \sum_{\lambda} \vec{v}_\lambda\vec{v}_\lambda^\dagger \mathbf{\cal S}^\dagger \right)
\end{equation}
but using the completeness relation, the sum over $\lambda$ yields just 1, and we have 
\begin{equation}
\mathrm{ tr}\mathbf{\cal S}\mathbf{\cal S}^\dagger = \mathrm{tr} \, \mathbf{\cal N}  = 1
\end{equation}
as we'd expect.

\end{document}